# Favoritism in Research Assistantship Selection in Turkish Academia

Dr. Osman Gulseven

Skyline University College

*Abstract*— This article analyzes the procedure for the initial employment of research assistants in Turkish universities to see if it complies with the rules and regulations. We manually collected 2409 applicant data from 53 Turkish universities to see. Our aim is to see if applicants are ranked according to the rules suggested by the Higher Education Council of Turkey. The rulebook states that applicants should be ranked according to a final score based on the weighted average of their GPA (30%), graduate examination score (30%), academic examination score (30%), and foreign language skills score (10%). Thus, the research assistant selection is supposed to be a fair process where each applicant is evaluated based on objective metrics. However, our analysis of data suggests that the final score of the applicants is almost entirely based on the highly subjective academic examination conducted by the hiring institution. Thus, the applicants' GPA, standardized graduate examination score, standardized foreign language score are irrelevant in the selection process, making it a very unfair process based on favoritism.

## 1. Introduction

According to the Turkish Higher Education Law number 2547-33, a research assistant (RA) is defined as those whose primary duties are research, investigation, help with experiments, and other tasks given by relevant departments. RAs are appointed for a maximum of three years and can be reappointed as long as they are in good academic status. Thanks to the low responsibility of staying as a research assistant, most graduate students are reluctant to finish their studies because graduation means losing their relaxed lifestyle. Once they graduate, they need to find a job that pays slightly higher. There is only a negligible difference (10% or less) between the salary of an assistant professor and a research assistant. However, being an assistant professor comes with additional responsibilities such as teaching exhaustive hours to overcrowded classes with hundreds of registered students.

The primary condition for being employed as a research assistant is to be a graduate student. You also need to enter the academic graduate student examination (ALES), which is a centralized Turkish version of GRE/GMAT exams. There is a minimum GPA requirement which depends on the faculty policy. There is also a foreign language requirement score of which is determined by the YOKDIL exam or equivalent score from similar exams. Finally, each department conducts a highly subjective academic exam (AE) to all applicants in the form of a written exam supposedly based on the departmental curriculum. Based on the Law of Higher Education 2547-50/d, the official guidelines set by the Higher Education Council dictate that the selection of RAs is based on the following semi-centralized procedure:

First, the hiring academic institutions announce their decision to employ new RAs by advertising these positions in the Official Gazette. The applicants send their application package directly to the hiring institution. It is a precondition to be of less than 35 years old and be a graduate student in good academic status. Any applicant should receive a minimum score of 70 out of 100 from the postgraduate studies examination (ALES). Another minimum requirement is to receive a grade of 50 or higher from a foreign language exam (YOKDIL or similar). Each institution has the right to increase the minimum scored for application. By law, it is mandatory to make the process as fair and as open possible. Thus, the law states explicitly that all institutions follow these guidelines to be equal opportunity employers.

Once the hiring institution receives all applications within a given deadline, the applicants are ranked based on their exam results (60% of ALES and 40% of YOKDIL scores). After this first round evaluation, a minimum of 10 applications for each open position is invited for an academic examination. This examination is not centralized and each academic department is responsible for conducting its own exam. Although the law explicitly states that the exam should be relevant to the subject matter, objective, measurable, and auditable in its contents, in most cases its highly subjective where academic corruption is rampant.

The final score of applicants is based on the following formula:

Final Score = 30% ALES + 30% undergraduate GPA + 30% Academic Exam + 10% YOKDIL

If the GPA is based on a 4-point scale, it is transformed into a 100-point scale using a formula determined by the Higher Education Council. This scaling ranges from 60 to 100 (i.e., GPA of 1 is converted as 60, and a GPA of 4 is converted as 100). The applicant who receives the highest score based on the above formula is formally invited for an employment contract. The second highest score is declared as the backup and is employed only if the highest scored applicant does not comply with the employment formalities on time.

In this article, we test whether the RA selection is a fair process that follows the guidelines suggested by the authorities. Specifically, we are testing whether the final score is indeed based on the weighted average of individual scores (30% ALES, 30% undergraduate GPA, 30% Academic Exam, and 10% YOKDIL). The results of 428 academic exams are investigated using the hiring data from 53 academic institutions. Specifically, we are looking to see whether the people who are offered employment contracts have higher academic status as measured by objective metrics, or whether they are chosen purely based on their score on academic examination results.

## 2. Literature Review

GPA is the primary indicator of students' academic self-efficacy and motivation for achievement (French, Immekus, & Oakes, 2005; Robbins et al., 2004). Thus, high school grades are very indicators of university grades with almost one to one correlation. Even further, high school GPA and retention in the university are highly correlated (Rohr, 2012). Extending that logic to higher education, one would expect the university GPA to be reflective of the graduate education GPA. GPA is also believed to be a strong indicator of the abilities and productivity of the student. Some employers even put a minimum threshold on undergraduate GPA for new recruits (Sulastri, Handoko, & Janssens, 2015).

Factors affecting the employee selection process has been a controversial issue for a long time. While this process is expected to be fair and objective, there are a variety of sociodemographic factors such as age, gender, affective in hiring (Haefner, 1977). In academia, female workers may find it harder to quit their jobs or switch to better, potentially higher-paying positions (Reagan & Maynard, 1974). Female and ethnic minority academicians are widely believed to have experience both explicit and implicit discrimination (Browne & Misra, 2003; Milkman, Akinola, & Chugh, 2012). Women in academia might face several challenges (Gulseven & Mostert, 2019; Mostert & Gulseven, 2019). Recent research on France suggests that those with French names are more likely to call for an interview compared to those with non-French sounding names (Edo, Jacquemet, & Yannelis, 2019).

One of the primary causes of systematic discrimination is the recruitment methodology employed by universities (Pruitt & Isaac, 2006). Most universities use informal recruitment channels such as "old boy network" and alumni connections with limited use of advertising media. While it is more cost-efficient to rely on such informal networks, the candidates come from a limited pool and the system is inherently biased towards the university's own alumnus. The role of attractiveness is also a fascinating subject. Some research suggests that the hiring managers pay attention to the attractiveness of the applicant, putting the less attractive female applicants at a systematic disadvantage (Marlowe, Schneider, & Nelson, 1996).

Only a few studies investigated the process of graduate student selection at selected universities to test the efficiency of the procedure (Arapgirlioglu, Zahal, Gurpinar, & Ozhan, 2014; Koğar, Sayin, & Assist, 2014; Öztürk & Anil, 2012). The common point in those studies is that the admission of students in the graduate education system in Turkey is unfair, where the admission interview is the primary determinant. We extend this study to the RA hiring process in Turkey, where the results also show that the RA selection process is also unfair.

## 3. Data

One of the reasons why there has not been any academic research conducted is the lack of data. A central database that includes the individual results at the individual selection level does not exist. Occasionally, some university departments become headlines due to the irregularities observed in their RA employee selections. This is when the RA selection data is well-publicized only for that specific academic examination result. Thanks to the law, it is also possible to find the same data at the webpages of hiring units. The data can be found in different forms: some departments release only the names of the hired RAs; some departments release only the names of the hired RAs and reserve RAs; some departments release both the names and detailed scores of hired RAs and reserved RAs. The form of data also shows wide diversity. While some departments post individual scores in original Excel form with detailed calculations, many departments post them in pdf form making it harder to retrieve.

Given the wide variety in the quality and form of data, the only way left to collect reliable data is by manual collection. Thus, each data is collected manually from the websites of relevant institutions. The 2019 Spring semester class of ECON 106 students helped with the data collection process. The initial database had a total of 2977 observations. However, 568 observations were deleted as these applicants did not take the academic examination, although they were invited to do so. The final dataset includes data on 2409 applicants who applied for the RA positions in 423 departments in 53 universities. 711 of those applicants applied for RA positions in engineering faculties.

Each entry consists of the name, faculty, and department of the hiring institution, the name, surname, and gender of the applicant, the scores received from ALES, undergraduate GPA, YOKDIL, and academic exam, and the rank of the applicant among those who applied for the same RA position. In terms of gender, the data is well balanced.

**Table 1. Applicants based on their final status and gender**

|  | Frequency | | |
|---|---|---|---|
| **Status** | **Female** | **Male** | **Total** |
| Employed | 272 | 258 | 530 |
| Fail | 764 | 649 | 1413 |
| Reserve | 243 | 223 | 466 |
| **Total** | **1279** | **1130** | **2409** |

As can be seen in Table 1, out of 2409 applicants, only 530 are offered employment contracts (About 22%). While that number might seem very good, we need to consider that these are only reported numbers. Thus, among those who passed the first stage, the ratio of employed is about 10%. The data on those fails are missing due to either they do not attend the academic exam, or their score is not reported by the relevant institution.

**Table 2. Mean and Standard Z scores of applicants**

| **Mean Values** | | | | | |
|---|---|---|---|---|---|
| **Result** | **ALES** | **YOKDIL** | **GPA** | **AE** | **Final Score** |
| **Employ** | 81.98 | 80.12 | 81.37 | 70.16 | 78.01 |
| **Fail** | 81.47 | 78.04 | 77.67 | 33.65 | 65.59 |
| **Reserve** | 82.21 | 79.91 | 81.76 | 56.86 | 74.17 |
| **TOTAL** | 81.72 | 78.86 | 79.27 | 46.18 | 69.98 |
| **Z-Values** | | | | | |
| **Result** | **ALES** | **YOKDIL** | **GPA** | **AE** | **Final Score** |
| **Employ** | 0.09 | 0.19 | 0.22 | 1.15 | 1.21 |
| **Fail** | -0.05 | -0.13 | -0.17 | -0.49 | -0.53 |
| **Reserve** | 0.06 | 0.18 | 0.26 | 0.18 | 0.23 |
| **TOTAL** | 0 | 0 | 0 | 0 | 0 |

Table 2 shows the average scores of those who are employed, failed, and in the reserve list. The mean scores were almost the same as the median scores, so only mean scores are reported. There is a highly visible difference in the mean academic examination score between those who are employed (70.16), Fail (33.65), Reserve (56.86) failed. While the mean ALES, YOKDIL, and GPA scores are close, the employed applicants have higher average academic exam scores, which inflates their final score. Thus, the difference between those employed and failed is much more visible in the academic exam category.

When the scores of those applicants in the fail and reserve list are compared, there are contradictory results. While one might expect that the applicants with higher GPA averages are more likely to be offered employment contracts, this is not the case. On the contrary, those in the reserve list have higher

ALES and GPA scores than those in the employment list. YOKDIL scores of employed are negligibly higher than that of reserved. What is striking is that the outcome of the so-called academic examination is the primary factor that determines the allocation of applicants into different lists.

The Z-scores within each examination are also calculated as each examination is separate, and the applicants are competing only with those who attend the same exam. The calculation of Z-scores is simply based on the difference between individual and examination-specific mean scores divided by the examination-specific standard deviation. Thus for applicant j, entering the examination i, the score is calculated according to the following formula:

$$Z\_Score_{ij} = \frac{X_{ij} - Mean(X_i)}{Std.Dev(X_i)} \quad (1)$$

When we measure the results in terms of standardized Z-Scores we observe almost the same outcomes. Those in the fail list have the lowest Z-scores in all dimensions. However, the employed applicants are not the top GPA earners. The average normalized GPA score of those in the reserve list is higher than the applicants who are offered employment. It is the academic examination results that determine who is employed and who is in the reserve list.

### 4. Results

Table 3 below lists which factors are correlated with the final score of the applicant. If it was a fair evaluation based on objective metrics, the correlation of ALES (30%), GPA (30%), YOKDIL (10%), and Academic Exam (30%) would be similar to their weights in score calculation. However, that is not the case, as observed below:

**Table 3. Correlation between employment factors**

| | ALES | GPA | YD | AE | Score |
|---|---|---|---|---|---|
| *Nominal Scores* | | | | | |
| **ALES** | 1.00 | | | | |
| **GPA** | 0.05 | 1.00 | | | |
| **YD** | 0.29 | 0.14 | 1.00 | | |
| **AE** | 0.01 | 0.13 | 0.11 | 1.00 | |
| **Score** | 0.27 | 0.45 | 0.34 | 0.90 | 1.00 |
| **Standardized Scores** | ALES_Z | GPA_Z | YD_Z | AE_Z | Score_Z |
| **ALES_Z** | 1.00 | | | | |
| **GPA_Z** | -0.07 | 1.00 | | | |
| **YD_Z** | -0.07 | 0.05 | 1.00 | | |
| **AE _Z** | -0.02 | 0.06 | 0.13 | 1.00 | |
| **Score_Z** | 0.11 | 0.37 | 0.22 | 0.87 | 1.00 |

The correlation analysis suggests that what the final score of the applicant is almost perfectly correlated with the result of the academic exam. The correlation between the final score and academic exam is 0.90 when measured in terms of nominal scores, and 0.87 when measured in terms of standardized scores. There is a moderate correlation between GPA and final score (0.45 and 0.37). ALES and YOKDIL scores seem relevant when they are measured in nominal terms, but this effect disappears when performance is measured in standardized scores. Thus the correlations with the final score can be ranked as follows: Subjective Academic Exam > Objective GPA > Objective YOKDIL > Objective ALES.

The discriminant analysis has been performed using both nominal and standardized values. The linear model suggests that there is no significant difference between the two datasets, so only the standardized results are reported here. The linear model was able to classify 67% of the data into true groups. The correct classification rate of 84.5% is much higher when estimating those employed. The model correctly estimated 70.9% of the fails whereas that rate is only 35% when categorizing those on the reserve list.

**Table 4. Linear Discrimination**

|  | Employed | Fail | Reserve |
|---|---|---|---|
| Constant | -1.2765 | -0.2540 | -0.0904 |
| ALES_Z | 0.2472 | -0.1329 | 0.1211 |
| GPA_Z | 0.3713 | -0.2381 | 0.2992 |
| YOKDIL_Z | 0.1286 | -0.1060 | 0.1759 |
| Academic Exam_Z | 2.1059 | -0.9074 | 0.3510 |

The linear discrimination factorial analysis in Table 4 above suggests that the score of the academic exam is distinctively highest among all groups, followed by GPA. YOKDIL score has the least effect on the outcome of who is employed and who fails the application, whereas ALES has the least effect on the outcome of the reserve list. Thus, whether an applicant is hired or not depends almost completely on what score s/he received in the subjective academic exam.

## 5. Conclusion

In this article, we discussed whether the employment of RAs is a fair and open process using statistical methods on sample data collected from individual hiring institutions in Turkey. According to the law, the applicants are ranked based on their final score, which is calculated as the weighted average of ALES, GPA, YOKDIL, and Academic Examination scores. By law, the weight of ALES, GPA, and Academic Examination is 30%, whereas YOKDIL has a weight of 10%. The hiring institution has no power on the calculation of ALES, GPA, and YOKDIL scores. However, the hiring committee prepares the questions for the academic examination and grades them according to applicants' performance. This is where it is claimed that the hiring committee abuses its authority to determine who is hired and who is not. Thus, it is widely believed that the fair and mechanical process of ranking the applicants is flawed due to the magnified role of academic examination.

In the linear discrimination model, the role of the academic exam is found to be distinctively higher than any other factor. The ordinal logistic regression has also been performed to test which factors affect the probability of someone being in the employment, reserve, or fail list. The results suggested that the ALES, GPA, and YOKDIL scores are not significant at all. It is purely the outcome of an academic exam that determines who is employed and who is in reserve or fail list.

The statistical data analysis employed here leaves no doubt that the RA selection system in Turkey is inherently flawed. While in theory, it sounds like a fair process where the final rank is based on a mechanical formula, in practice, this is not the case. The results suggest that whether someone is employed or not depends purely on the highly subjective academic examination score which conducted and evaluated by the hiring department. It might be argued that the exam is fair and there is no evidence regarding the fairness of the exam. However, the data suggest a negative correlation between an applicant's GPA and academic examination. The aim of the academic examination is to test the knowledge of the applicant in the subject matter. This knowledge is best reflected in the GPA of the applicant. A student takes about 40 to 50 courses during his/her studies. If each course has 3 exams, that means about 120 to 150 exams excluding the project work, quizzes, home works, presentations, etc. The outcome of these exams is reflected in the student's GPA. Interestingly, the role GPA when employing an RA is almost negligible due to the dominance of the so-called academic examination.